\documentstyle[preprint,eqsecnum,aps]{revtex}
\newcommand{\be}{\begin{equation}}
\newcommand{\ee}{\end{equation}}
\newcommand{\bea}{\begin{eqnarray}}
\newcommand{\eea}{\end{eqnarray}}
\begin{document}
\draft
\title{\hbox{\hspace{20em} \normalsize  PUPT-1554, 1995} \Large
\bf{3-D Perturbations in Conformal Turbulence}}

\author{ L. Moriconi$^{\dag}$}
\address{ Instituto de F\'\i sica, Universidade Federal do Rio de Janeiro
\\C.P. 68528, Rio de Janeiro, RJ --- 21945-970, Brasil}
\maketitle

\begin{abstract}
The effects of three-dimensional perturbations in two-dimensional turbulence
are investigated, through a conformal field theory approach.
We compute scaling exponents for the energy spectra of enstrophy and energy
cascades, in a strong coupling limit, and compare them to the values found
in recent experiments.
The extension of unperturbed conformal turbulence to the present
situation is performed by means of a simple physical picture in which the
existence of small scale random forces is closely related to deviations of the
exact two-dimensional fluid motion.
\end{abstract}


\newpage
\section{Introduction}
It has been recognized that turbulence, with its manifold experimental
realizations, is one of the challenging problems
to which very different
approaches, ranging from pure mathematics to engineering applications,
have been developed in an interesting complementary way.
One of the most important methods to study turbulence is, in fact,
the formulation via field theory, based on its relationship with stochastic
partial differential equations \cite{domi,yak}.
However, such technique is far from being
well established and complete, so that new ideas and important improvements
are constantly appearing in the subject.

Recently, Polyakov suggested that
non-unitary minimal models of conformal field theory could be used to describe
the physics of two-dimensional turbulence \cite{poly}.
The advantage of this proposal is that one can deal in a controllable way
with a set of anomalous dimensions and short distance products.
An infinite number of inertial range exponents follows from this approach
\cite{lowe,falko,matsuo}
and one of the still open problems is how to find
``selection rules," which would define the experimentally relevant
minimal models or the connection between them and
statistical characterizations of the random forces acting on the system.
These ideas have attracted the attention of many
authors, and generalizations have
been investigated, as, for instance, possible boundary effects \cite{chung},
alternative physical pictures for the enstrophy and energy cascades
\cite{cateau}, 
and magnetohydrodynamic turbulence \cite{ferri}.

We will consider, in this paper, the problem of conformal turbulence
including in its formalism the influence of three-dimensional effects.
Our motivation comes from a number of experimental studies, in
which approximately two-dimensional fluids were observed, together with the
unavoidable presence of three-dimensional perturbations
\cite{hopfi,mory1,mory2,max}.
It was verified that a quasi two-dimensional fluid is perturbed by small scale
forces originated from the the degrees of freedom related to the direction
perpendicular to the plane of motion.
We will take this fact into account, noticing that there are
also compressibility effects which cannot be neglected in an effective
two-dimensional theory of the perturbed system. 
A generalization of the conformal approach will be devised and new
inertial range exponents will be obtained here, in reasonable agreement
with the experimental data.

This paper is organized as follows. In the next section we briefly review
the most important and practical aspects of conformal turbulence, in order
to make the paper as self contained as possible. In section III, we discuss
some of the experiments carried out to investigate two-dimensional
turbulence. This will motivate us to define an effective (and infinite)
set of stochastic partial differential equations which represents a quasi
two-dimensional fluid under the influence of three-dimensional perturbations.
The conformal approach is then introduced in order to solve the Hopf equations
for the turbulence problem. Furthermore, the constant enstrophy and energy
flux conditions are also studied. Explicit solutions are found and described
in section IV and, in section V, the problem of boundary effects is discussed.
Finally, in section VI we comment on our results and on possible
directions for future investigations.

\section{Conformal Turbulence}
The minimal models of conformal field theory \cite{bpz} are
generically defined by a
pair of relatively prime numbers, $(p,q)$, with $p<q$. These models contain
a subset of
$(p-1)(q-1)/2$ scalar primary operators, $\psi_{(m,n)}$,
labelled by $1\leq m <p$
and $1 \leq n \leq (q-1)/2$, if $p$ is even, or
$1\leq m \leq (p-1)/2$ and $1 \leq n <q$, otherwise, having
dimensions $\Delta_{(m,n)}=\left( (pn-qm)^2-
(p-q)^2 \right)/4pq$. The reason for the choice of scalar operators is that
we will be dealing with isotropic correlation functions in the
turbulence problem. The operator product expansion (OPE) of two primary
operators $\psi_{(r_1,s_1)}(z)$ and $\psi_{(r_2,s_2)}(z')$,
with $|z-z'| \rightarrow 0$ is written as
\bea
&&\psi_{(r_1,s_1)}(z)\psi_{(r_2,s_2)}(z')=
\sum_{(r_3,s_3)} (a \bar a)^{ \left( \Delta_{(r_3,s_3)}
-\Delta_{(r_1,s_1)}-\Delta_{(r_2,s_2)} \right)}
\sum_{(n,m)} C^{(r_3,s_3)}_{ \{ (n_1,...,n_k);
(m_1,...,m_l) \} } \nonumber \\
&&\times
L_{-n_1}...L_{-n_k} \bar L_{-m_1}...\bar L_{-m_l}
a^{\sum n} \bar a^{\sum m} \psi_{(r_3,s_3)}(z) \ , \ \label{ct1}
\eea
where $|r_1-r_2|+1 \leq r_3 \leq \hbox{min}(r_1+r_2-1,2p-r_1-r_2-1)$,
$|s_1-s_2|+1 \leq s_3 \leq \hbox{min}(s_1+s_2-1,2q-s_1-s_2-1)$ and we have
introduced, in (\ref{ct1}), the Virasoro generators of conformal
transformations, $L_{-n}$ and $\bar L_{-n}$.
The interest in these models is related not only to their finite number
of primary operators, but also to the fact that their dimensions and the form of
short distance products are completely known.

Let us look now at the problem of turbulence in two dimensions and show how it
may be matched \cite{poly} with the above operator structures.
The motion of an
incompressible fluid is assumed, even in the turbulent regime, to be described
by the Navier-Stokes equations for the velocity field,
\be
\partial_t v_\alpha + \left( \delta_{\alpha \gamma} - {{\partial_\alpha
\partial_\gamma} \over {\partial^2}} \right) v_\beta \partial_\beta v_\gamma
= f_\alpha + \nu \partial^2 v_\alpha \ , \ \label{ct2}
\ee 
where $f_\alpha$ represents a random force acting at large scales, determined
by a characteristic length $L$, and $\nu
\rightarrow 0$ is
the viscosity, associated to the small scale where dissipation effects
come into play, yielding a natural UV cutoff for the system.
In terms of the stream function, $\psi$, related to the velocity
field by $v_\alpha = \epsilon_{\beta \alpha} \partial_\beta \psi$, we may write
the following equation for the vorticity field, $\omega = \partial^2 \psi$,
\be
\partial_t \omega + \epsilon_{\alpha \beta} \partial_\alpha \psi
\partial^2 \partial_\beta \psi = \epsilon_{\alpha \beta} \partial_\alpha
f_\beta + \nu \partial^2 \omega \ . \ \label{ct3}
\ee
One of the fundamental problems of turbulence theory is to find solutions of
the Hopf equations, for statistical averages over realizations of the velocity
field,
\be
\partial_t [ <\omega(x_1,t) \omega(x_2,t)...\omega(x_n,t)> ] = 0 \ , \
\label{ct4}
\ee
where the time derivative is expressed through the use of equations (\ref{ct3}).
In the inertial range, the standard view of the problem is that both forcing
and viscosity terms may be neglected in order to formulate an effective set
of Hopf equations. Considering, furthermore, the convection term in (\ref{ct3})
as a vanishing point-splitted product of fields, that is, $\oint_{|z-z'|=|a|}
(dz' / a) \epsilon_{\alpha \beta} \partial_\alpha \psi(z)
\partial^2 \partial_\beta \psi(z') \rightarrow 0$, when $|z-z'| \rightarrow 0$,
we would have, then, an exact solution of (\ref{ct4}). A concrete
realization of this possibility
may be achieved if we regard the stream function $\psi$ as
a primary operator of some conformal minimal model. In this case we may use
all the available information on operator dimensions and OPE's to
extract physical results from the analysis of the problem.
According to this assumption, let $\phi$ be the primary operator which has
the lowest dimension, $\Delta \phi$, appearing in the OPE $\psi \psi$,
between fields with the same dimension $\Delta \psi$. Taking $a \equiv
|a| \exp(i \theta)$, we will have, thus,
\bea
&& \lim_{|a| \rightarrow 0} \oint_{|z-z'|=|a|} {{dz'} \over a}
\epsilon_{\alpha \beta}
\partial_\alpha \psi(z) \partial^2 \partial_\beta \psi(z') \nonumber \\
&&\sim \int d \theta \left [ \partial^2_{\bar a} \partial_a \partial_z
-\partial^2_a \partial_{\bar a} \partial_{\bar z} \right ]
(a \bar a)^{\left( \Delta \phi - 2 \Delta \psi \right)}
\sum C_{\{n;m\}}L_{-n_1}...L_{-n_k} \bar L_{-m_1}...\bar L_{-m_l}
a^{\sum n} \bar a^{\sum m} \phi(z, \bar z) \nonumber \\
&&\sim (a \bar a)^{\left( \Delta \phi - 2 \Delta \psi \right)}
\left [ L_{-2} \bar L_{-1}^2 - \bar L_{-2} L_{-1}^2 \right ] \phi
\ , \  \label{ct5}
\eea 
as the dominant contribution in this short distance product. It is important to
note that in order to get (\ref{ct5}) it was necessary to set $C_{\{1;2\}}=
C_{\{2;1\}}$ and $C_{\{1;(1,1)\}}=C_{\{(1,1);1\}}$,
as it follows from the pseudoscalar nature
of the $\epsilon$ factor above. We see, then, that (\ref{ct5}) vanishes with
$|a| \rightarrow 0$ if
\be
\Delta \phi > 2 \Delta \psi \ , \ \label{ct6}
\ee
which is one of the constraints that the chosen minimal model has to satisfy.
An additional constraint comes from the condition of a constant
enstrophy or energy flux through the inertial range, meaning that $<\dot \omega
(x) \omega(0)> \sim x^0$ or $<\dot v_\alpha(x) v_\alpha(0)> \sim x^0$,
respectively. In the case of a constant enstrophy flux, we have
\be
<\dot \omega(x) \omega(0)> \sim
(a \bar a)^{\left( \Delta \phi - 2 \Delta \psi \right)}
<\left [ \left(  L_{-2} \bar L_{-1}^2 - \bar L_{-2} L_{-1}^2 \right)
\phi(x) \right ] \partial^2 \psi(0)> \ . \ \label{ct7}
\ee 
The correlation function at the RHS of (\ref{ct7}) is now evaluated by means
of a purely dimensional argument, as $L^{-2 \left( \Delta \phi + \Delta \psi
+3 \right)}$, which makes sense if one thinks that there is an effective
IR cutoff in the theory at the scales where the forcing terms act. Imposing
(\ref{ct7}) to be independent of $L$, we get
\be
\Delta \phi + \Delta \psi + 3 = 0 \ . \ \label{ct8}
\ee
In the case of an energy cascade, the argument is the same and the constraint
turns out to be $\Delta \phi + \Delta \psi + 2 = 0$. It is known that there is
an infinite number of minimal models compatible with (\ref{ct6}) and
(\ref{ct8}) \cite{lowe}. The general belief, and still
an open problem, is that there
may be universality classes, associated to the statistical properties of the
forcing terms, which would single out one or another of the possible solutions.

An alternative analysis of conformal turbulence regards the existence of
boundary effects on the vacuum expecation values (VEV's) of single operators
in non-unitary theories \cite{zamo}. In this case, one has to consider the OPE
between $\phi(x)$ and $\psi(0)$ in (\ref{ct7}), picking up the most relevant
operator, let us say, $\chi$. Now, (\ref{ct8}) is modified to
$\Delta \phi + \Delta \psi -\Delta \chi +3 = 0$, with an analogous change
for the constant energy flux condition. Some of these further solutions
(in the enstrophy cascade picture)
were obtained in ref. \cite{chung}.

The connection of the conformal approach with real experiments or numerical
simulations is made
through the computation of inertial range exponents,
which describe the decrease of energy in
the region of higher Fourier modes. In the situation where VEV's of single
operators vanish, the inertial range exponents are given by $4 \Delta \psi
+1$ and, in the opposite case, by $4 \Delta \psi -2\Delta \phi +1$.
A good agreement has been reached between the former possibility, for the
the direct enstrophy cascade case, and numerical simulations \cite{legras,babiano} of the two-dimensional Navier-Stokes equations.

\section{Three-dimensional effects}
In a series of interesting experiments, Hopfinger et al.
\cite{hopfi,mory1,mory2} studied
the turbulence phenomenon as it happens in a rotating tank, where at its 
bottom there was an oscilating grid responsible for perturbations of the
fluid motion. According to the Taylor-Proudman theorem
\cite{proudman,taylor,inglis} a rotating
fluid tends to behave as if it were two-dimensional and in fact this was
observed in the form of coherent structures
(vortices) organized in the direction parallel to the rotation axis of
the tank. However, ``defects" in the vortices were seen to propagate from
the very turbulent region at the bottom of the tank up to the effectively
two-dimensional system. The essential picture extracted from these observations
is that the fluid should be best described in terms of two-dimensional equations
containing not only large scale forcing terms but also small scale random 
perturbations, originated from either vortex-breakdown or soliton pulses
propagating along
vorticity filaments.
The experimental data suggested then the existence of an inertial range,
likely to be related to a direct enstrophy cascade
and well approximated by $E(k) \sim k^{-2.5}$,
which represents a less steep energy spectrum
than the one obtained by Kraichnan \cite{kraich}, $E(k) \sim
k^{-3}$, or even other proposals \cite{saff,moff}, not excluding
conformal turbulence
\cite{lowe}. This puzzingly result is presently understood to be due only to the
measurement techniques used in the experiments, based on the analysis of the
dispersion of suspended particles in the fluid \cite{mory2}.
More recently, similar
experiments were conduced by Narimousa et al. \cite{max} and direct measurements
of the turbulent velocity field were recorded. The results pointed out the
existence of a possible energy spectra $E(k) \sim k^{-5/3}$ at lower
wavenumbers, in agreement with the conjecture of an inverse energy cascade
\cite{kraich},
and a range at higher wavenumbers, where $E(k) \sim k^{-5.5 \pm 0.5}$. In
this region, the spectral slope was seen to depend on the controlling
external conditions, with results varying from $E(k) \sim k^{-5.0}$ up to
$E(k) \sim k^{-6.0}$. It is worth to note that a spectral law
$E(k) \sim k^{-5}$ follows from Rhines theory of $\beta$-plane turbulence
\cite{rhines} and, alternatively, is closely approximated by some solutions
of the constant enstrophy flux condition in the conformal approach, like
the minimal models $(9,71)$ or $(11,87)$.

The variation of exponents obtained in the experiments may have a theoretical
counterpart in the existence of a set of operator anomalous dimensions,
making it interesting to analyze the problem from the conformal field theory
point of view. It is clear, however, that the inertial range exponents, found in
ref. \cite{lowe}, cannot reproduce the experimental situation. We believe that the important ingredient, missing in the previous conformal approach, is precisely the existence of three-dimensional perturbations, which must be taken into account in any realistic model of a quasi two-dimensional fluid.

In view of the above considerations, let us write the two-dimensional
Navier-Stokes equations as
\be
\partial_t v_\alpha + v_\beta \partial_\beta v_\alpha =
\nu \partial^2 v_\alpha + f^{(1)}_\alpha + gf^{(2)}_\alpha -\partial_\alpha P
\ , \ \label{tde1}
\ee
where $f^{(1)}_\alpha$ and $f^{(2)}_\alpha$ are stirring forces defined at
large $(L)$ and small $( \mu << L)$ scales, respectively. The dimensionless
constant $g$ represents,
roughly, a coupling with the three-dimensional modes of the fluid.
We assume that the dissipation scale, $a$, is related , in principle, to the
other scales of the problem as $a<< \mu << L$.
This means that even though the perturbations act at very small scales, when 
compared to the macroscopic size of the system, they are still much larger
than the scale where dissipation occurs.

An important point here is that the condition of incompressibility, when
formulated in three dimensions, reads $\partial_1 v_1 +
\partial_2 v_2 +\partial_3 v_3 = 0$, 
suggesting that the ``projection" of this
constraint
to the two-dimensional
world has to be given by $\partial_\alpha v_\alpha = {\cal O} (g)$,
in the framework of
equations (\ref{tde1}). The velocity field may be described, then,
by means of a stream function, $\psi$, and a velocity potential, $\phi$, as
\be
v_\alpha = \epsilon_{\beta \alpha} \partial_\beta \psi + g \partial_\alpha
\phi \ . \ \label{tde2}
\ee
It is of further interest to study, besides the vorticity $\omega$,
the divergence of $v_\alpha$,
given by $\rho = g \partial^2 \phi$. An exact, although infinite, chain  of
equations may be generated if we expand $\psi$ and $\phi$ in powers
of $g$, substituing them into (\ref{tde1}) and collecting the coefficients
of the obtained series. Defining, in this way,
\bea
&&\psi = \sum^\infty_{n=0} g^n \psi^{(n)} \ , \
\omega= \sum^\infty_{n=0} g^n \omega^{(n)} \ , \ \nonumber \\
&&\phi = \sum^\infty_{n=0} g^n \phi^{(n)} \ , \ 
\rho= \sum^\infty_{n=0} g^{n+1} \rho^{(n)} \ , \
\label{tde3}
\eea
we get the following set of coupled equations,
\bea
\hbox{{\it i) }} \partial_t \omega^{(n)} &&+ \sum^n_{p=0} \epsilon_{\alpha \beta}
\partial_\alpha \psi^{(p)}\partial_\beta \partial^2 \psi^{(n-p)} +
\sum^{n-1}_{p=0} \left[ \partial_\beta \phi^{(p)} \partial_\beta
\partial^2 \psi^{(n-p-1)} +
\partial^2 \phi^{(p)} \partial^2 \psi^{(n-p-1)} \right] \nonumber \\
&&= \nu \partial^2 \omega^{(n)} + \epsilon_{\alpha \beta} \partial_{\alpha}
f^{(2)}_\beta \delta_{n,1} \ , \
\nonumber \\
\hbox{{\it ii) }} \partial_t \omega^{(0)} &&+ \epsilon_{\alpha \beta}
\partial_\alpha \psi^{(0)}\partial_\beta \partial^2 \psi^{(0)} =
\nu \partial^2 \omega^{(0)}+\epsilon_{\alpha \beta} \partial_{\alpha} 
f^{(1)}_\beta \ , \ \nonumber \\
\hbox{{\it iii) }} \partial_t \rho^{(n)} &&+ \sum^{n-1}_{p=0}
\left[ \partial_\alpha \partial_\beta \phi^{(p)}
\partial_\alpha \partial_\beta \phi^{(n-p-1)} +
\partial_\alpha \phi^{(p)} \partial_\alpha \partial^2 \phi^{(n-p-1)}
\right] \nonumber \\ 
&&+ \sum^n_{p=0} \left[2 \epsilon_{\alpha \beta}
\partial_\beta \partial_\sigma \phi^{(p)}
\partial_\alpha \partial_\sigma \psi^{(n-p)} 
+\epsilon_{\alpha \beta} \partial_\alpha \psi^{(n-p)} \partial_\beta
\partial^2\phi^{(p)}  \right] \nonumber \\
&&+\sum^{n+1}_{p=0} \left[ \partial_\alpha \partial_\beta \psi^{(p)}
\partial_\alpha \partial_\beta \psi^{(n-p+1)}
- \partial^2 \psi^{(p)} \partial^2 \psi^{(n-p+1)} \right]
= \nu \partial^2 \rho^{(n)} \ , \
\nonumber \\ 
\hbox{{\it iv) }} \partial_t \rho^{(0)} &&+2\partial_\alpha \partial_\beta
\psi^{(0)}\partial_\alpha \partial_\beta \psi^{(1)}
+2\epsilon_{\alpha \beta}\partial_\beta \partial_\sigma \phi^{(0)}
\partial_\alpha \partial_\sigma \psi^{(0)}+
\epsilon_{\alpha \beta} \partial_\alpha \psi^{(0)} \partial_\beta
\partial^2\phi^{(0)} \nonumber \\
&&-2 \partial^2 \psi^{(0)} \partial^2 \psi^{(1)}
=\nu \partial^2 \rho^{(0)}+\partial_\alpha f^{(2)}_\alpha
\ , \ \label{tde4}
\eea
and, finally, the constraint of incompressibility for the
$g$-independent part of the velocity field, which defines the pressure term,
\be
\left( \partial_\alpha \partial_\beta \psi^{(0)} \right)
\left( \partial_\alpha \partial_\beta \psi^{(0)} \right) - \partial^2
\psi^{(0)} \partial^2 \psi^{(0)} = \partial_\alpha f^{(1)}_\alpha -
\partial^2 P \ . \ \label{tde4.5}
\ee 
In the above expressions, $n \geq 1$. We have obtained, therefore, a set
of stochastic partial differential equations.
In a statistical
description, reflecting a stable asymptotic limit for the correlation
functions of $\omega$ and $\rho$, Hopf equations
may be straightforwardly written as
\be
\partial_t < \prod_{i=1}^{N} \omega^{(n_i)}(x_i,t) \prod_{j=N+1}^{M}
\rho^{(n_j)}(x_j,t) > = 0 \ . \ \label{tde5}
\ee
We observe now that in (\ref{tde4}), equation $ii)$ is identical
to the one which corresponds to an unperturbed ($g=0$) two-dimensional fluid.
This means that the field $\psi^{(0)}$ will be related to an enstrophy 
or energy cascade, even in the presence of three-dimensional effects. This
field plays the role of an external random variable
in the other equations, since its dynamics is independent of the other
components $\psi^{(n)}$ or to the field $\phi$
(in general, the subset $\{ \psi^{(0)}, \psi^{(1)},...,\psi^{(n)},
\phi^{(0)},\phi^{(1)},...,\phi^{(n-1)} \}$ contains fields which act like
external random perturbations in the equations for $\psi^{(p)}$ and
$\phi^{(p-1)}$, with $p \geq n+1$).
Considering that (\ref{tde4}) gives relatively complex equations, the
analysis of the problem might seem hopeless, being perhaps adressed only to
a numerical treatment. However, we can extend the conformal approach, applied
previously to the unperturbed case, to find here solutions of the Hopf
equations. Our basic assumption is that not only $\psi^{(0)}$ but
also the other components in the power expansions of $\psi$ and $\phi$
are primary operators which belong to some minimal model in a conformal field
theory. It is necessary, then, to define a scale $\ell$, possibly associated to
intermittency effects, which allows us to write the following dimensionally
correct expansion,
\bea
&&\psi = \sum_{n=0}^\infty f_n \ell^{2 (\Delta \psi^{(n)} - \Delta \psi^{(0)})}
g^n \psi^{(n)} \ , \ \nonumber \\
&&\phi = \sum_{n=0}^\infty f_n' \ell^{2 (\Delta \phi^{(n)} - \Delta \psi^{(0)})}
g^n \phi^{(n)} \ , \ \label{tde6} 
\eea
where $\psi^{(n)}$ and $\phi^{(n)}$ have dimensions $\Delta \psi^{(n)}$ and
$\Delta \phi^{(n)}$, respectively.

The introduction of a scale $\ell$ in
(\ref{tde6}) means that the perturbed system exhibits a breaking of scale invariance in the inertial range. We will see later that this phenomenon is
signaled by the existence of constant enstrophy or energy fluxes which
depend on the small scales of three-dimensional perturbations. It is conceptually important to understand the physical origin of $\ell$. A clue for this comes from the structure of couplings between $\psi^{(0)}$ and the other fields, as expressed in equation (\ref{tde4}). As we have already observed,
$\psi^{(0)}$ is effectively an external field in the equations for $\psi^{(n)}$
(with $n \geq 1$) and $\phi^{(n)}$ (for any $n$). In this way, it is plausible to have a relation between $\ell$ and the scales involved in the dynamics of
$\psi^{(0)}$. Now, if we consider the turbulent limit of the equations for
$\psi^{(0)}$, corresponding to $\nu \rightarrow 0$ (or, alternatively,
$a \rightarrow 0$), we are left essentially with the correlation length $L$ of
large scale random forces. A simple choice, thus, is to consider $\ell = L$.
In this respect, one may observe that the small scale $\mu$ could also be used in the definition of $\ell$. We have, however, physical reasons to believe that this does not happen: $\mu$ is related to the forcing terms in the equations for $\psi^{(1)}$ and $\phi^{(1)}$, which we expect to be irrelevant when compared to the nonlinear convection terms in the range of wavenumbers given by $|\vec k| << 1/ \mu$.

It is interesting to note that there is an analogy between our problem and the statistical mechanics of second order phase transitions for a system close to its critical point. In this case, one can study deviations of the critical temperature $T_c$ by means of an expansion in $(T-T_c)$ and through the use of the operator structure of the critical theory \cite{amit}. Here, in the turbulence context, the ``critical theory" is just what we get when $g \rightarrow 0$.

In order to simplify the notation we will keep using (\ref{tde3}), with the above observations in mind. We are interested to obtain possible
combinations of primary operators, in eq. (\ref{tde6}),
that would not affect, in the limit $\mu \rightarrow 0$, the constant enstrophy or energy fluxes, obtained from the dynamics of the field $\psi^{(0)}$.
Within this point of view,  
it is important, therefore, to consider short distance products of a certain
number operators, as it follows from (\ref{tde4}). Taking two
generic primary operators, $O^{(p)}_1$ and $O^{(p')}_2$
(for example, $\phi^{(p)}$ and $\psi^{(p')}$),
with dimensions $\Delta O^{(p)}_1$ and $\Delta O^{(p')}_2$,
respectively, we may write
\bea
&&O^{(p)}_1(z,\bar z) O^{(p')}_2(z', \bar z')
= (a \bar a)^{\left(
\Delta A^{(p,p')}_{O_1,O_2} - \Delta O^{(p)}_1 - \Delta O^{(p')}_2 \right)}
\nonumber \\
&&\times \sum C^{O_1^{(p)},O_2^{(p')}}_{\{(n_1,...,n_k);(m_1,...,m_\ell) \}}
L_{-n_1}...L_{-n_k} \bar L_{-m_j}...\bar L_{-m_\ell}
a^{\sum n} \bar a^{\sum m} A^{(p,p')}_{O_1,O_2}(z,\bar z) \ , \
\label{tde7}
\eea
where $A^{(p,p')}_{O_1,O_2}$ is the primary operator with the
lowest dimension in the above OPE. The short distance products
appearing in (\ref{tde4}) are listed below
together with the conformal field
theory representation, obtained after straightforward computations:
\bea
&&\bullet \epsilon_{\alpha \beta} \partial_\alpha
\psi^{(p)}\partial_\beta \partial^2 \psi^{(p')}
\sim (a \bar a)^{\left (\Delta A^{(p,p')}_{\psi \psi}- \Delta \psi^{(p)}
-\Delta \psi^{(p')} \right)}
\left[ L_{-2} \bar L_{-1}^2 - \bar L_{-2} L_{-1}^2 \right]
A^{(p,p')}_{\psi \psi}
\nonumber \\
&&\bullet \partial_\beta \phi^{(p)} \partial_\beta \partial^2 \psi^{(p')} 
\sim \partial^2 \phi^{(p)} \partial^2 \psi^{(p')} \sim  
(a \bar a)^{\left (\Delta A^{(p,p')}_{\phi \psi}- \Delta \phi^{(p)}
-\Delta \psi^{(p')} -1\right)}
L_{-1} \bar L_{-1} A^{(p,p')}_{\phi \psi} \nonumber \\
&&\bullet \partial_\alpha \partial_\beta \phi^{(p)}
\partial_\alpha \partial_\beta \phi^{(p')}
\sim \partial_\alpha \phi^{(p)} \partial_\alpha \partial^2 \phi^{(p')} \sim 
(a \bar a)^{\left (\Delta A^{(p,p')}_{\phi \phi}- \Delta \phi^{(p)}
-\Delta \phi^{(p')} -1 \right)}
L_{-1} \bar  L_{-1} A^{(p,p')}_{\phi \phi} \nonumber \\ 
&&\bullet  \epsilon_{\alpha \beta} \partial_\beta \partial_\sigma
\phi^{(p)} \partial_\alpha \partial_\sigma \psi^{(p')}
\sim \epsilon_{\alpha \beta} \partial_\alpha
\psi^{(p')}\partial_\beta \partial^2 \phi^{(p)} \sim 
(a \bar a)^{\left (\Delta A^{(p,p')}_{\phi \psi}- \Delta \phi^{(p)}
-\Delta \psi^{(p')} \right)} \nonumber \\
&&\times \left[ L_{-2} \bar L_{-1}^2 - \bar L_{-2} L_{-1}^2 \right]
A^{(p,p')}_{\phi \psi}
\nonumber \\
&&\bullet \partial_\alpha \partial_\beta \psi^{(p)}
\partial_\alpha \partial_\beta \psi^{(p')} \sim
\partial^2 \psi^{(p)} \partial^2 \psi^{(p')} \sim
(a \bar a)^{\left (\Delta A^{(p,p')}_{\psi \psi}- \Delta \psi^{(p)}
-\Delta \psi^{(p')} -1 \right)} A^{(p,p')}_{\psi \psi}
 \ , \ \label{tde8} 
\eea
with $p,p' \geq 0$, except in the last relation, for the product
of the type $\psi \psi$, where $p+p'\geq 1$.
From (\ref{tde8}) we see clearly that Hopf equations are satisfied if
\bea
&&\hbox{{\it i)} } \Delta A^{(0,0)}_{\psi \psi}- 2 \Delta \psi^{(0)}
> 0 \ , \ \nonumber \\
&&\hbox{{\it ii)} } \Delta A^{(p,p')}_{\psi \psi}- \Delta \psi^{(p)}
-\Delta \psi^{(p')}-1> 0 \ , \ \nonumber \\
&&\hbox{{\it iii)} } \Delta A^{(p,p')}_{\psi \phi}- \Delta \psi^{(p)}
-\Delta \phi^{(p')}-1> 0 \ , \ \nonumber \\
&&\hbox{{\it iv)} } \Delta A^{(p,p')}_{\phi \phi}- \Delta \phi^{(p)}
-\Delta \phi^{(p')}-1> 0 \ , \ \label{tde9}
\eea
with $p+p' \geq 1$ in {\it ii)} and $p,p' \geq 0$ in {\it iii)}
and {\it iv)}. These equations
are the first step in the generalization
of unperturbed conformal turbulence, in order to deal with a larger set
of primary operators. We will find more inequalities, restricting, then, up
to some extent the number of different operators allowed in the theory. 
Let us now write the conditions for constant enstrophy or energy fluxes
through the inertial range. Here we will assume that vacuum expectation
values of primary operators are zero.
The alternative possibility is discussed in the next section.
The case of a constant enstrophy flux requires,
as commented before, that we compute $<\dot \omega(x) \omega(0)>$.
From relations (\ref{tde4}) we get
\bea
&&<\dot \omega (x) \omega (0)> =
\nonumber \\
&&\sum_{n,m}^\infty \sum_{p=0}^n g^{n+m} < \left \{ 
\epsilon_{\alpha \beta}
\partial_\alpha \psi^{(p)} \partial_\beta \partial^2 \psi^{(n-p)}
+g \left[ \partial_\beta
\phi^{(p)} \partial_\beta
\left. \partial^2 \psi^{(n-p)} + \partial^2 \phi^{(p)} \partial^2 \psi^{(n-p)}
\right] \right \} \right \vert_x \nonumber \\
&&\times \partial^2 \psi^{(m)}(0)> \ . \ \label{tde10} 
\eea
In the above expression we may define a ``large scale" part as the one
which depends solely on the field $ \psi^{(0)}$ and a ``small scale" part,
involving the fields $\phi^{(p)}$ and the other components of $\psi$.
Regarding the large scale part, we already know, from the study of the
unperturbed case, that the constant enstrophy flux condition is
\be
\Delta A^{(0,0)}_{\psi \psi} + \Delta \psi^{(0)}
+ 3 = 0 \ . \ \label{tde11.1}
\ee
It is natural to assume, like in the case of unperturbed
conformal turbulence, that the correlation functions in the small scale
part may also be evaluated by means of a dimensional argument, where,
instead of using the typical large scale
parameter $L$, the correct choice turns out to be the small length scale $\mu$.
Assuming, furthermore, that $\mu \rightarrow 0$ leads to a well defined
limit, we just require the powers of $\mu$ in the most relevant terms
belonging to the small scale part of (\ref{tde10}) (contributions which have
the lowest power of $a \bar a$) to be non-negative numbers.
This discussion may be restated by saying that we will need to select one
or more of the following conditions,
\bea
&&\hbox{{\it i)} } \Delta A^{(p,p')}_{\phi \psi} + \Delta \psi^{(p'')}
+ 2 \leq 0 \ , \ \nonumber \\
&&\hbox{{\it ii)} } \Delta A^{(p,p')}_{\psi \psi} + \Delta \psi^{(p'')}
+ 3 \leq 0 \ , \ \label{tde11}
\eea 
according to the analysis of the dominant terms in the small scale part of
$<\dot \omega(x) \omega(0)>$.
In the derivation of (\ref{tde11}) we have used the OPE's computed from
the Hopf equations, given by (\ref{tde8}).
An additional care must be taken if it happens that
$A^{(p,p')}_{\psi \psi} = \psi^{(p'')}$ or
$A^{(p,p')}_{\phi \psi}=\psi^{(p'')}$ for some values of $p,p'$ and $p''$.
In this circumstance it is necessary to have 
$\Delta \psi^{(p'')} = -3/2$ or $\Delta \psi^{(p'')} = -1$, respectively,
to assure spatially independent correlation functions and hence a constant
enstrophy flux. 

Let us turn now to the case of a constant energy flux. We have
\bea
&&< \dot v_\alpha(x) v_\alpha(0)> = \nonumber \\
&&- \sum_{n,m}^\infty \sum_{p=0}^n
g^{n+m} < \left \{ g^2 \partial_\beta \phi^{(p)} \partial_\beta
\partial_\alpha \phi^{(n-p)} + g \left[ \epsilon_{\gamma
\alpha} \partial_\beta \phi^{(p)} \partial_\beta \partial_\gamma \psi^{(n-p)}
\right.
+\epsilon_{\sigma \beta}
\partial_\sigma \psi^{(p)} \partial_\beta 
\partial_\alpha \phi^{(n-p)} \right] \nonumber \\
&&\left. \left. +\partial_\alpha \partial^{-2} \left[
\partial^2 \psi^{(0)} \partial^2 \psi^{(0)}
-\partial_\sigma \partial_\beta \psi^{(0)}
\partial_\sigma \partial_\beta \psi^{(0)} \right] \delta_{n,0}
+ \epsilon_{\sigma \beta} \epsilon_{\gamma \alpha}
\partial_\sigma \psi^{(p)} \partial_\beta
\partial_\gamma \psi^{(n-p)}
\right \}  \right \vert_x \nonumber \\
&&\times \left(
\epsilon_{\eta \alpha} \partial_\eta \psi^{(m)}(0)+g \partial_\alpha \phi^{(m)}
(0) \right)> \ . \ 
\label{tde12}
\eea
Here we cannot refer immediately to the Hopf equations and formulate
a set of conditions, as we did in the constant enstrophy flux case.
There are, in (\ref{tde12}), OPE's which do not appear in (\ref{tde4}), viz.
\bea
&&\bullet \epsilon_{\sigma \beta} \epsilon_{\gamma z}
\partial_\sigma \psi^{(p)} \partial_\beta \partial_\gamma \psi^{(p')}
\sim  (a \bar a)^{\left (\Delta A^{(p,p')}_{\psi \psi}- \Delta \psi^{(p)}
-\Delta \psi^{(p')} -1 \right)} L_{-1} A^{(p,p')}_{\psi \psi} \nonumber \\
&&\bullet \epsilon_{\sigma \beta} \epsilon_{\gamma z}
\partial_\sigma \psi^{(0)} \partial_\beta \partial_\gamma \psi^{(0)}
+\partial_z \partial^{-2} \left[
\partial^2 \psi^{(0)} \partial^2 \psi^{(0)}
-\partial_\sigma \partial_\beta \psi^{(0)}
\partial_\sigma \partial_\beta \psi^{(0)} \right] \nonumber \\
&&\sim (a \bar a)^{ \left( \Delta A^{(0,0)}_{\psi \psi} -2 \Delta \psi^{(0)}
\right) } 
\left[ \bar L_{-1} L_{-2} -4 \partial^{-2} L_{-1}^3 \bar L_{-2} \right]
A^{(0,0)}_{\psi \psi} \nonumber \\
&&\bullet \epsilon_{\gamma z} \partial_\beta \phi^{(p)} \partial_\beta
\partial_\gamma \psi^{(p')} \sim
\epsilon_{\sigma \beta} \partial_\sigma \psi^{(p')} \partial_\beta
\partial_z \phi^{(p)} \sim
(a \bar a)^{\left (\Delta A^{(p,p')}_{\phi \psi}- \Delta \phi^{(p)}
-\Delta \psi^{(p')} -1\right)} L_{-1} A^{(p,p')}_{\phi \psi}
\nonumber \\
&&\bullet \partial_\beta \phi^{(p)} \partial_\beta \partial_z \phi^{(p')}
\sim (a \bar a)^{\left (\Delta A^{(p,p')}_{\phi \phi}- \Delta \phi^{(p)}
-\Delta \phi^{(p')} -1\right)} L_{-1} A^{(p,p')}_{\phi \phi}
\ . \ \label{tde13}
\eea
In (\ref{tde13}) we have $\epsilon_{\gamma z} \equiv (
\epsilon_{\gamma 1} - i \epsilon_{\gamma 2} )/2$.
Using the above point-splitted products, we can get the constant energy
flux conditions.
First, the equation following from the large scale part of $< \dot v(x) v(0)>$, 
\be
\Delta A^{(0,0)}_{\psi \psi} + \Delta \psi^{(0)} +2 =0
\label{tde14.1}
\ee
and then the inequalities which come from the small scale terms,
\bea
&&\hbox{{\it i) }} \Delta A^{(p,p')}_{\psi \psi} + \Delta \psi^{(p'')} +1
+ \delta_{p+p',0} \leq 0
\ , \ \nonumber \\
&&\hbox{{\it ii) }} \Delta A^{(p,p')}_{\psi \psi} + \Delta \phi^{(p'')} +1
+ \delta_{p+p',0} \leq 0
\ , \ \nonumber \\
&&\hbox{{\it iii) }} \Delta A^{(p,p')}_{\phi \phi}+ \Delta \psi^{(p'')}+1
\leq 0
\ , \ \nonumber \\
&&\hbox{{\it iv) }} \Delta A^{(p,p')}_{\phi \phi}+ \Delta \phi^{(p'')}+1
\leq 0
\ , \ \nonumber \\
&&\hbox{{\it v) }} \Delta A^{(p,p')}_{\phi \psi}+ \Delta \phi^{(p'')}+1
\leq 0
\ , \ \nonumber \\
&&\hbox{{\it vi) }} \Delta A^{(p,p')}_{\phi \psi}+ \Delta \psi^{(p'')}+1
\leq 0
\ , \ \label{tde14}
\eea 
where, analogously to the enstrophy cascade case, only a subset
of (\ref{tde14}) has to be considered.
Supplementary relations, like the ones obtained after equation (\ref{tde11}),
are in order, to avoid possible $x$-dependent correlation functions.
We are led, here, to
\bea
&&\hbox{{\it i) }} \Delta \psi^{(p)} = -1 \hbox{, if }
A_{\psi \psi}^{(0,0)} = \psi^{(p)} \ , \ \nonumber \\
&&\hbox{{\it ii) }} \Delta \phi^{(p)} = -1 \hbox{, if }
A_{\psi \psi}^{(0,0)} = \phi^{(p)} \ , \ \nonumber \\
&&\hbox{{\it iii) }} \Delta \psi^{(p'')} = -1/2 \hbox{, if }
A_{\psi \psi}^{(p,p')} = \psi^{(p'')} \hbox{, for $p+p'>0$, or }
A_{\phi \psi}^{(p,p')} = \psi^{(p'')} \nonumber \\
&&\hbox{ or }
A_{\phi \phi}^{(p,p')} = \psi^{(p'')} \ , \ \nonumber \\
&&\hbox{{\it iv) }} \Delta \phi^{(p'')} = -1/2 \hbox{, if }
A_{\psi \psi}^{(p,p')} = \phi^{(p'')} \hbox{, for $p+p'>0$, or }
A_{\phi \psi}^{(p,p')} = \phi^{(p'')} \nonumber \\
&&\hbox{ or }
A_{\phi \phi}^{(p,p')} = \phi^{(p'')} \ . \ \label{tde15}
\eea

Once we have some solution at hand, derived from the conditions
obtained here, we may associate inertial range exponents to each one
of the fields $\psi^{(p)}$ and $\phi^{(p)}$, expressed by $4 \Delta
\psi^{(p)}+1$ and $4 \Delta \phi^{(p)}+1$, respectively.
From these values, we have to select the one which will appear effectively
in experimental situations. This problem is investigated in the
following section.

\section{analysis of the constant flux conditions}
We have obtained so far all the conditions necessary to find
minimal models related to an enstrophy or energy cascade in
a quasi two-dimensional fluid. In order to explore them, the first
observation we can make is that these models
must belong to the infinite set of
solutions found in the former study of unperturbed conformal turbulence.
This follows directly from the conditions which depend only on
$\psi^{(0)}$.
A strategy of computation could be, thus, just a numerical analysis of all
possible combinations of fields for these previously known minimal models.
As straight it may sound, this approach is hardly useful when the number of
primary operators becomes large, a fact that happens already for the first few
minimal models.

A more interesting computational scheme is provided if we look for solutions of
the form
\bea
&& \psi = \psi_0 + f_a(g)\psi_1 \ , \ \nonumber \\
&& \phi = f_b(g) \phi_0 \ , \ \label{afc1}
\eea
where $f_a(0)=f_b(0)=0$, that is, we are considering solutions with
$\psi^{(p)} = \psi_1$, for $p \geq 1$, and $\phi^{(p)} = \phi_0$, for any $p$. 
This approach is valuable since a little reflection shows that if 
it is impossible to satisfy the constant flux conditions through any pair
of fields $\psi_1$ and $\phi_0$, then there are no further solutions
for the model under consideration.
All our task is, therefore, to consider the set of minimal models representing
conformal turbulence without perturbations, from which the fields $\psi_0$
may be immediately obtained, and add, according to the new constraints
associated to three-dimensional effects, the fields $\psi_1$ and $\phi_0$.

In the study of the inertial range exponents, we may think of, at least, three
limits for $f_{a,b}(g)$: a) $ g \rightarrow 0$, that is, $f_{a,b}(g)
\rightarrow 0$, b) $f_{a,b}(g) \simeq 1$, and c) $g >>1$, which may be defined
as a ``strong coupling" regime. In the first case, the perturbations play a
negligible role and everything is described by unperturbed conformal turbulence.
A competition between exponents shows up in the second case, where the less
steep spectral slope will be the most relevant in the limit of higher
wavenumbers. We see, in this way, that cases a) and b) cannot give any of
the steeper spectral slopes observed in real experiments. The third case is,
in fact, where we have some hope to find a relation with experimental results.
It would be unphysical to have $f_{a,b}(g) \rightarrow 0$, for large values of
$g$, since in this limit we would recover the unperturbed system.
Also, it is unlikely to have $f_{a,b}(g) \rightarrow$ {\it const.}: taking, for instance, gaussian random forces $f^{(1)}_\alpha$ and $f^{(2)}_\alpha$, with $<f^{(1)}_\alpha (\vec x,t) f^{(2)}_\beta (\vec x',t')>=0$ and $<f^{(2)}_\alpha (\vec x,t) f^{(2)}_\beta (\vec x',t')>= D_{\alpha \beta} (|\vec x - \vec x'|) \delta (t-t')$, it may be proved, from the retarded nature of the diffusion propagator, that
$<f^{(2)}_\alpha (\vec x,t) v_\beta (\vec x',t)> = g D_{\alpha \beta} (|\vec x - \vec x'|)$, yielding
\be
{\partial \over {\partial g} } \left [
{ {<f^{(2)}_\alpha (\vec x,t) v_\beta (\vec x',t)>} \over
{D_{\alpha \beta} (|\vec x - \vec x'|)}} \right ] = 1 \ . \ \label{afc2}
\ee
Let us assume, thus, that $f_{a,b}(g)$ diverges as $g \rightarrow \infty$.
This means that the inertial range exponent derived from $\psi_0$ may be
discarded and we have to analyze only the competition between the exponents
obtained from $\psi_1$ and $\phi_0$.

We performed an investigation of the first six minimal models for both the
enstrophy and energy cascade cases. In the enstrophy case we found solutions
for all the models studied. They are represented in table 1 and in
figure 1. In table 1, we show the fields $\psi_1$ and $\phi_0$ 
for the minimal models (2,21), (3,25) and (3,26), together with their
associated inertial range exponents. As the number of solutions became larger,
we had to represent the other three models, (6,55), (7,62) and (8,67) in
figure 1, where we plotted the most relevant exponents, found from the
competition between $\psi_1$ and $\phi_0$, in the strong coupling regime.
We observe, from the results, that there is a good agreement with experimental
verifications, with the only considerable deviation occuring for the very small
set of two solutions for the model (2,21).
The solutions, excluding the model (2,21), were organized in table 3, where
values of mean exponents and standard deviations are described. It is clearly
seen that the perturbed exponents are in general lesser than the exponents
of the unperturbed fluid.

In the energy case, an interesting fact happened: most of the models studied
did not yield any solution for the fields $\psi_1$ and $\phi_0$. Only the
model (10,59), represented in table 2, gave solutions, all of them
with inertial range exponents close to $-3.0$, which do not support
the conjecture of a Kolmogorov exponent, $-5/3$, for the range of lower
wavenumbers. However, more theoretical and experimental work is necessary in
order to arrive at a conclusive answer on this point. 

\section{boundary perturbations}
It is worth to understand what happens when boundary effects are supposed to
have some influence in the problem of conformal turbulence. Below, we obtain
the set of conditions needed to account for it, leaving a numerical
analysis for future investigations.

The basic modification here is that we have to study further OPE's in the
conditions of constant enstrophy or energy fluxes, found in section III,
since now VEV's of single operators do not necessarily vanish. In this
way, let us define the primary operator
$A_{\left( O_1O_2 \right) O_3}^{(p,p')p''}$ as the one with the
lowest dimension appearing in the OPE
$\left( O_1^{(p)} O_2^{(p')} \right) O_3^{(p'')}$, where the product of
$O_1^{(p)}$ and $O_2^{(p)}$ was computed first. The conditions we are looking
for must be obtained from the analysis of the $x$-dependence of the dominant
terms in (\ref{tde10}) and (\ref{tde12}). In the situation of a constant
enstrophy flux, the large scale part of (\ref{tde10}) gives
\be
\Delta A^{(0,0)}_{\psi \psi} + \Delta \psi^{(0)}
-\Delta A_{\left( \psi \psi \right) \psi }^{(0,0)0}+ 3 = 0 
\ , \ \label{bp1}
\ee
which is nothing else than the condition established in section II, in a
diferent notation. On the other side, the small scale part of (\ref{tde10})
gives one or more of the following conditions:
\bea
&&\hbox{{\it i)} } \Delta A^{(p,p')}_{\phi \psi} + \Delta \psi^{(p'')}
-\Delta A_{\left( \phi \psi \right) \psi}^{(p,p')p''}
+ 2 = 0 \ , \ \nonumber \\
&&\hbox{{\it ii)} } \Delta A^{(p,p')}_{\psi \psi} + \Delta \psi^{(p'')}
- \Delta A_{\left( \psi \psi \right) \psi}^{(p,p')p''}
+ 3 = 0 \ . \ \label{bp2}
\eea 
A similar analysis for the case of an energy cascade yields, for the large and
small scale parts of (\ref{tde12}), respectively,
\be
\Delta A^{(0,0)}_{\psi \psi}
+ \Delta \psi^{(0)}
- \Delta A_{\left( \psi \psi \right) \psi}^{(0,0)0}+2 =0
\label{bp3}
\ee
and
\bea
&&\hbox{{\it i) }} \Delta A^{(p,p')}_{\psi \psi} + \Delta \psi^{(p'')}
- \Delta A_{\left( \psi \psi \right) \psi}^{(p,p')p''}
+1 + \delta_{p+p',0} = 0
\ , \ \nonumber \\
&&\hbox{{\it ii) }} \Delta A^{(p,p')}_{\psi \psi} + \Delta \phi^{(p'')}
- \Delta A_{\left( \psi \psi \right) \phi}^{(p,p')p''}
+1 + \delta_{p+p',0} = 0
\ , \ \nonumber \\
&&\hbox{{\it iii) }} \Delta A^{(p,p')}_{\phi \phi}+ \Delta \psi^{(p'')}
-  \Delta A_{\left( \phi \phi \right) \psi}^{(p,p')p''} + 1 = 0
\ , \ \nonumber \\
&&\hbox{{\it iv) }} \Delta A^{(p,p')}_{\phi \phi}+ \Delta \phi^{(p'')}
- \Delta A_{\left( \phi \phi \right) \phi}^{(p,p')p''} + 1 = 0
\ , \ \nonumber \\
&&\hbox{{\it v) }} \Delta A^{(p,p')}_{\phi \psi}+ \Delta \phi^{(p'')}
- \Delta A_{\left( \phi \psi \right) \phi}^{(p,p')p''} + 1 = 0
\ , \ \nonumber \\
&&\hbox{{\it vi) }} \Delta A^{(p,p')}_{\phi \psi}+ \Delta \psi^{(p'')}
- \Delta A_{\left( \phi \psi \right) \psi}^{(p,p')p''} + 1 = 0
\ . \ \label{bp4}
\eea
The computation of inertial range exponents is also modified. We now have to
consider all possible combinations like $\psi^{(p)} \psi^{(p')}$ and
$\phi^{(p)} \phi^{(p')}$ in the evaluation of the velocity-velocity
correlation function. The observed inertial range exponent
must be obtained from 
$2 \Delta \psi^{(p)} + 2 \Delta \psi^{(p')} - 2 \Delta A_{\psi \psi}^{(p,p')}
+1$ or $2 \Delta \phi^{(p)} + 2 \Delta \phi^{(p')}
- 2 \Delta A_{\phi \phi}^{(p,p')}+1$.

\section{conclusion}
The problem of two-dimensional turbulence was investigated, taking into account
the presence of three-dimensional perturbations. They were introduced in
an effective way, represented by random forcing terms which act at small
scales in the two-dimensional Navier-Stokes equations, as suggested by 
experimental observations. A coupling constant, $g$, related to
the strength of these
additional forces, allowed us to write a power expansion for the velocity
field, containing also a compressible part. An infinite set
of equations was found by just collecting terms with the same powers of $g$.
The components $\psi^{(p)}$ and $\phi^{(p)}$, appearing in the power expansion
of the velocity field were assumed to be primary operators of
some conformal minimal model. We obtained, then, from
point-splitted products of operators, a group of conditions in order to have a
solution of the Hopf equations and to reproduce the situation of a constant
enstrophy or energy flux through the inertial range.
In the constant flux conditions, large and small scale terms were defined and
evaluated by means of an extension of the dimensional argument employed
formely in the study of analogous correlation functions.
An analysis of the first six minimal models of unperturbed conformal turbulence
was performed, showing that the picture of a constant enstrophy cascade is in
good agreement with experimental data, yielding inertial range exponents, for
the strong coupling regime, $g>>1$, very close to the ones observed in the
laboratory. Regarding the energy cascade case, we noticed that most of the
minimal models considered in our study were unable to give solutions for the
perturbed system. Only one solution was obtained, with inertial range exponents
around $-3.0$. It would be interesting to investigate further minimal models
for the energy case, in order to see if a closer connection with the results
indicated in experiments could be reached.

From tables 1 and 2, and figure 1, we see clearly that there are many solutions,
differing by just one of the fields $\psi_1$ or $\phi_0$, which give exactly
the same inertial range exponents, in the strong coupling regime. It is
tempting, then, to conjecture that one could find ``plateaux" for the
spectral slopes, while varying some set of external parameters. This question
is contained, of course, in the deeper problem of how to match large scale
properties of the fluid with the minimal models describing the inertial range.

A point which deserves attention is the crossover between unperturbed conformal turbulence and the results obtained in the strong coupling regime. A bridge between these two situations may be investigated not only by varying $g$, as we did in section IV, but also through $\mu \rightarrow 0$, when the effects of small scale three-dimensional perturbations on the constant enstrophy or energy fluxes become negligible. Finally, it is important to stress that a standard direct numerical simulation of equations (\ref{tde4})-(\ref{tde4.5}), up to some level $n$ in their hierarchy, would be an interesting way to study the above questions and the physical assumptions addressed in the present work.

\acknowledgments
I would like to thank C. Callan and D. Gross for the kind hospitality at the
Physics Department of Princeton University, where this paper was completed.
Also, I would like to thank A. Migdal for sharing his insights on the general
turbulence problem, and M. Moriconi for interesting comments. This work was
supported by CNPq (Brazil).

\newpage
\begin{center}
{\bf TABLE AND FIGURE CAPTIONS}
\end{center}
$ $

{\bf Table 1.}
Solutions for the constant enstrophy flux condition.
The first six models of unperturbed conformal turbulence were analyzed,
all of them yielding possible definitions of $\psi_1$ and $\phi_0$.
Here we show the results for the minimal models (2,21), (3,25) and (3,26).

{\bf Table 2.} 
Solutions for the constant energy flux condition.
The analysis of the first six models of unperturbed conformal
turbulence showed that most of them were ``blocked" by the presence of
perturbations. The only obtained solution corresponds to the model (10,59).

{\bf Table 3.}
Statistical data related to the solutions found for the constant enstrophy
flux condition, in the strong coupling regime, where a comparison
is made with the unperturbed values of the inertial range exponents.

{\bf Figure 1.}
Graphic representation of the inertial range exponents
in the enstrophy cascade case.
Figures a), b) and c) refer to the minimal models (6,55), (7,62) and (8,67),
respectively.
The horizontal axis, labelled by (m,n), represents the most relevant field
between $\psi_1$ and $\phi_0$, in the strong coupling regime.
The ordering of fields is the same as in the tables.

\newpage
{\bf Table 1.} 
\begin{center}
\begin{tabular}{|c|c|c|c|} \hline
\multicolumn{4}{|c|}{minimal model: (2,21)} \\ \hline
$\psi_1$ & $\phi_0$ & $4 \Delta \psi_1 +1$ & $4 \Delta \phi_0 +1$    \\ \hline
(1,9)  & (1,9)  & -7.38 & -7.38 \\ \hline
(1,9)  & (1,10) & -7.38 & -7.57 \\ \hline\hline
\multicolumn{4}{|c|}{minimal model: (3,25)} \\ \hline
$\psi_1$ & $\phi_0$ & $4 \Delta \psi_1 +1$ & $4 \Delta \phi_0 +1$    \\ \hline
(1,6)  & (1,6) & -4.8 & -4.8 \\ \hline
(1,6) & (1,7) & -4.8 & -5.24 \\ \hline
(1,6) & (1,8) & -4.8 & -5.44 \\ \hline
(1,6) & (1,9) & -4.8 & -5.4 \\ \hline
(1,6) & (1,10) & -4.8 & -5.12 \\ \hline
(1,6) & (1,11) & -4.8 & -4.6 \\ \hline
(1,7) & (1,6) & -5.24 & -4.8 \\ \hline
(1,7) & (1,7) & -5.24 & -5.24 \\ \hline
(1,7) & (1,8) & -5.24 & -5.44 \\ \hline
(1,7) & (1,9) & -5.24 & -5.4 \\ \hline
(1,7) & (1,10) & -5.24 & -5.12 \\ \hline
(1,7) & (1,11) & -5.24 & -4.6 \\ \hline
(1,8) & (1,7) & -5.44 & -5.24 \\ \hline
(1,8) & (1,9) & -5.44 & -5.4 \\ \hline
(1,8) & (1,11) & -5.44 & -4.6 \\ \hline
(1,9) & (1,6) & -5.4 & -4.8 \\ \hline
(1,9) & (1,8) & -5.4 & -5.44 \\ \hline
(1,9) & (1,10) & -5.4 & -5.12 \\ \hline
\end{tabular}
\begin{tabular}{|c|c|c|c|} \hline
$\psi_1$ & $\phi_0$ & $4 \Delta \psi_1 +1$ & $4 \Delta \phi_0 +1$    \\ \hline
(1,10) & (1,6) & -5.12 & -4.8 \\ \hline
(1,10) & (1,7) & -5.12 & -5.24 \\ \hline
(1,10) & (1,8) & -5.12 & -5.44 \\ \hline
(1,10) & (1,9) & -5.12 & -5.4 \\ \hline
(1,10) & (1,10) & -5.12 & -5.12 \\ \hline
(1,10) & (1,11) & -5.12 & -4.6 \\ \hline
(1,11) & (1,6) & -4.6 & -4.8 \\ \hline
(1,11) & (1,7) & -4.6 & -5.24 \\ \hline
(1,11) & (1,8) & -4.6 & -5.44 \\ \hline
(1,11) & (1,9) & -4.6 & -5.4 \\ \hline
(1,11) & (1,10) & -4.6 & -5.12 \\ \hline
(1,11) & (1,11) & -4.6 & -4.6 \\ \hline\hline
\multicolumn{4}{|c|}{minimal model: (3,26)} \\ \hline
$\psi_1$ & $\phi_0$ & $4 \Delta \psi_1 +1$ & $4 \Delta \phi_0 +1$    \\ \hline
(1,6) & (1,6) & -4.96 & -4.96 \\ \hline
(1,6) & (1,7) & -4.96 & -5.46 \\ \hline
(1,6) & (1,8) & -4.96 & -5.73 \\ \hline
(1,6) & (1,9) & -4.96 & -5.77 \\ \hline 
(1,6) & (1,10) & -4.96 & -5.58 \\ \hline
(1,6) & (1,11) & -4.96 & -5.15 \\ \hline
(1,7) & (1,6) & -5.46 & -4.96 \\ \hline
(1,7) & (1,7) & -5.46 & -5.46 \\ \hline
(1,7) & (1,8) & -5.46 & -5.73 \\ \hline
\end{tabular}
\end{center}

\newpage
\begin{center}
\begin{tabular}{|c|c|c|c|} \hline
$\psi_1$ & $\phi_0$ & $4 \Delta \psi_1 +1$ & $4 \Delta \phi_0 +1$    \\ \hline
(1,7) & (1,9) & -5.46 & -5.77 \\ \hline
(1,7) & (1,10) & -5.46 & -5.58 \\ \hline
(1,7) & (1,11) & -5.46 & -5.15 \\ \hline
(1,8) & (1,7) & -5.73 & -5.46 \\ \hline
(1,8) & (1,9) & -5.73 & -5.77 \\ \hline
(1,8) & (1,11) & -5.73 & -5.15 \\ \hline
(1,9) & (1,6) & -5.77 & -4.96 \\ \hline
(1,9) & (1,8) & -5.77 & -5.73 \\ \hline
(1,9) & (1,10) & -5.77 & -5.58 \\ \hline
(1,10) & (1,6) & -5.58 & -4.96 \\ \hline
(1,10) & (1,7) & -5.58 & -5.46 \\ \hline
(1,10) & (1,8) & -5.58 & -5.73 \\ \hline
(1,10) & (1,9) & -5.58 & -5.77 \\ \hline
(1,10) & (1,10) & -5.58 & -5.58 \\ \hline
(1,10) & (1,11) & -5.58 & -5.15 \\ \hline
(1,11) & (1,6) & -5.15 & -4.96 \\ \hline
(1,11) & (1,7) & -5.15 & -5.46 \\ \hline
(1,11) & (1,8) & -5.15 & -5.73 \\ \hline
(1,11) & (1,9) & -5.15 & -5.77 \\ \hline
(1,11) & (1,10) & -5.15 & -5.58 \\ \hline
(1,11) & (1,11) & -5.15 & -5.15 \\ \hline
\end{tabular}
\end{center}

\newpage
{\bf Table 2.}
\begin{center}
\begin{tabular}{|c|c|c|c|} \hline
\multicolumn{4}{|c|}{minimal model: (10,59)} \\ \hline
$\psi_1$ & $\phi_0$ & $4 \Delta \psi_1 +1$ & $4 \Delta \phi_0 +1$    \\ \hline
(1,6) & (1,6) & -3.07 & -3.07 \\ \hline
(1,6) & (2,12) & -3.07 & -3.06 \\ \hline
(1,6) & (3,18) & -3.07 & -3.05 \\ \hline
(1,6) & (4,24) & -3.07 & -3.04 \\ \hline
(2,12) & (1,6) & -3.06 & -3.07 \\ \hline
(2,12) & (2,12) & -3.06 & -3.06 \\ \hline
(2,12) & (3,18) & -3.06 & -3.05 \\ \hline
(2,12) & (4,24) & -3.06 & -3.04 \\ \hline
(3,18) & (1,6) & -3.05 & -3.07 \\ \hline
(3,18) & (2,12) & -3.05 & -3.06 \\ \hline
(3,18) & (3,18) & -3.05 & -3.05 \\ \hline
(4,24) & (1,6) & -3.04 & -3.04 \\ \hline
(4,24) & (2,12) & -3.04 & -3.06 \\ \hline
(4,24) & (4,24) & -3.04 & -3.04 \\ \hline
\end{tabular}
\end{center}
\vspace{1 cm}

{\bf Table 3.}
\begin{center}
\begin{tabular}{|c|c|c|c|} \hline
minimal model & exponent (g=0) & mean exponent ($g \not = 0$) & standard
deviation \\ \hline
(3,25) & -4.6 & -4.90 & 0.28 \\ \hline
(3,26) & -4.23 & -5.25 & 0.27 \\ \hline
(6,55) & -3.73 & -5.89 & 0.21 \\ \hline
(7,62) & -4.03 & -5.46 & 0.28 \\ \hline
(8,67) & -4.51 & -4.90 & 0.34 \\ \hline
\end{tabular}
\end{center}

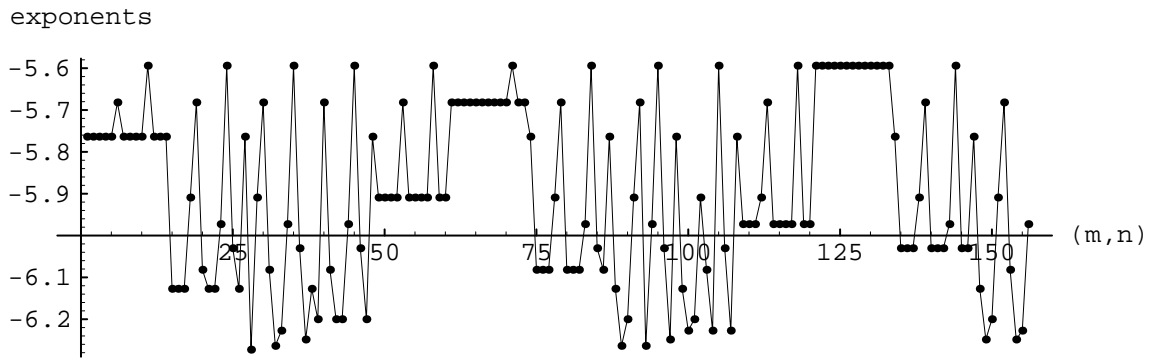
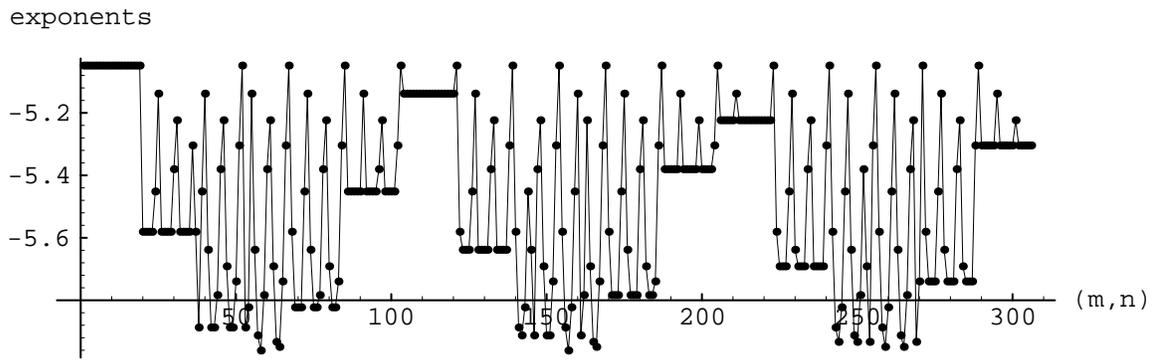
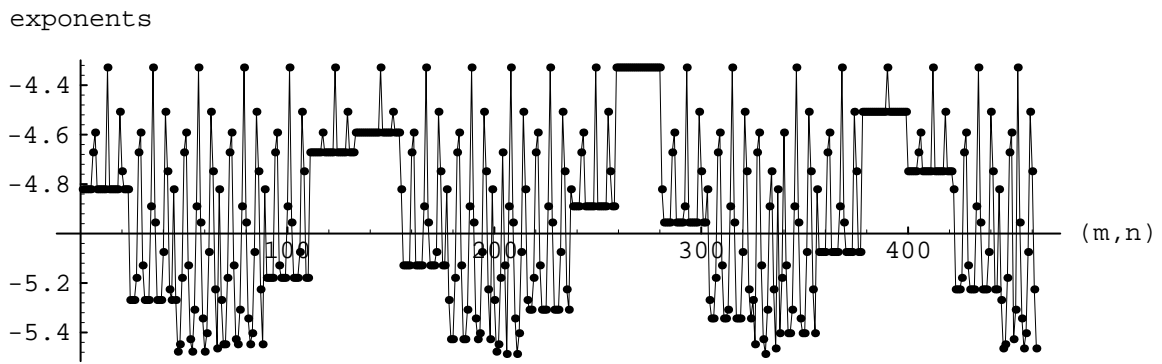
\end{document}